\documentclass[twocolumn,notitlepage,nofootinbib,amsmath,amssymb,aps,prd]{revtex4-1}

\usepackage[T1]{fontenc}
\usepackage{graphicx}
\usepackage{dcolumn}
\usepackage{bm}

\usepackage{color}
\usepackage[export]{adjustbox}

\begin{document}

\title{A one-stop function for gravitational-wave detection, identification and inference}

\author{Alvin J. K. Chua}
\email{alvincjk@nus.edu.sg}
\affiliation{Department of Physics, National University of Singapore, Singapore 117551}
\affiliation{Department of Mathematics, National University of Singapore, Singapore 119076}
\affiliation{Theoretical Astrophysics Group, California Institute of Technology, Pasadena, CA 91125, U.S.A.}

\date{\today}

\begin{abstract}
I define here a novel function on a modeled space of gravitational-wave signals, before studying its properties as a statistic for detection, as an objective function for identification, and as an effective likelihood function for inference. The main motivation behind this work is the open data-analysis problem for signals from extreme-mass-ratio inspirals, which is severely hindered by the presence of strong non-local parameter degeneracy in the signal space. I demonstrate the utility of the proposed function for the analysis of such signals, and suggest various possible directions for future research.
\end{abstract}

\maketitle


\section{Introduction}
\label{sec:intro}

The scientific inverse problem in gravitational-wave (GW) astronomy comprises three distinct data-analysis procedures (in addition of course to the vital tasks of noise characterization and forward source modeling, both of which we will assume can be achieved at full accuracy for present purposes). In the order they are performed, these procedures are:
\begin{itemize}
    \item Detection: Establishing the statistically significant presence of a GW signal in noisy detector data;
    \item Identification: Mapping the detected signal (sufficiently) accurately to the source parameters of a (sufficiently) representative forward model;
    \item Inference: Estimating the Bayesian posterior probability of the actual source parameters.
\end{itemize}
Detection can be performed with or without models, while identification requires at least an approximate model, and inference relies on a highly accurate one. The combination of detection and identification is commonly referred to as ``search''. For \emph{almost} all classes of GW source, identification is unnecessary when inference in the full model space is feasible, and/or guaranteed when accurate models are straightforward to invert (hence its complete neglect in much of the literature).

In the case of the extreme-mass-ratio inspirals (EMRIs) that will be observed by the near-future ESA--NASA mission LISA \cite{2017arXiv170200786A,Amaro-Seoane:2012lgq,Babak:2017tow,Berry:2019wgg}, the traditional dichotomy of detection and inference (or ``parameter estimation'') breaks down. Indeed, the overall EMRI inverse problem remains an open one, with well-documented obstacles arising from the theoretical challenges in achieving the required accuracy and extensiveness for models \cite{Barack:2018yvs,Pound:2021qin}; computational limitations in attaining sufficient model efficiency \cite{Chua:2020stf,Katz:2021yft}; as well as the inherent difficulty in exploring the voluminous EMRI signal space, due to \emph{strong and non-local parameter degeneracy}. This last feature was recently characterized qualitatively by Chua \& Cutler \cite{Chua:2021aah}, and is the primary hindrance to all three procedures of detection (uncertain criteria for reliable candidates), identification (multiple spurious candidates across signal space), and inference (impractical without highly localized priors) within the EMRI data-analysis problem. Past work on EMRI search \cite{Gair:2008ec,Babak:2009ua,MockLISADataChallengeTaskForce:2009wir,Cornish:2008zd,Wang:2012xh} has sought to provide practical solutions, but has not shown that identification in particular can be accomplished both reliably and efficiently.

For a fresh take on the problem, I propose in this manuscript a ``one-stop'' function with various favorable geometrical and statistical properties for EMRI detection, identification and inference. These properties are analytically and numerically characterized to an extent that should in my opinion leave little doubt over the utility of the function in realistic applications; stress tests on full-scale EMRI simulations are consequently left for future studies. The function takes the form
\begin{equation}\label{eq:function}
    f(\theta):=X(\theta)\exp{\left(-\frac{1}{2}\beta(\theta)\chi^2(\theta)\right)}.
\end{equation}
Here, $X$ is the standard matched-filtering detection statistic for a single signal template (Eq.\,\eqref{eq:detstat}); $\beta>0$ is a temperature-like calibration factor that is \emph{fully specified by the template model} (Eq.\,\eqref{eq:calibration}); and $\chi^2$ reduces for actual signals to a chi-squared statistic with $M$ degrees of freedom (Eq.\,\eqref{eq:chisq}), arising from a suitable decomposition of the template model into $M>1$ modes. Eq.\,\eqref{eq:function} is then to be used in a stochastic sampling algorithm, where the target probability density is $\propto\exp{f}$ (and so its logarithm, which is the commonly supplied quantity, is $f+\mathrm{const.}$). It has no tuning parameters.

The manuscript also touches on a larger second thesis: that most classes of coherent (phase-matching) statistics employed in stochastic GW searches are actually suboptimal due to \emph{uncontrolled variations} of the search statistic over the model space, caused by non-local signal correlations as well as the manifestation of detector noise. Such variations can generally be eliminated in $f$ for any specific application of coherent search, through the choice of $\chi^2$ (the mode decomposition) and $\beta$. While these are tailored here to the qualitative nature of the EMRI signal space, they might also be defined for example to suppress the impact of transient noise artifacts, or to aid future searches for long-duration signals from precessing and eccentric comparable-mass binaries (where non-local degeneracy could likewise be an issue).

\section{Derivation}
\label{sec:derivation}

In this section, we will expand on the derivation of Eq.\,\eqref{eq:function} in some detail. Our starting point is the generic decomposition of a GW-signal template $h(\theta)$ as
\begin{equation}\label{eq:modedecomp}
    h(\theta)=\sum_{m=1}^Mh_m(\theta)+\epsilon(\theta),
\end{equation}
with the following assumptions, valid for all $m,m'$:
\begin{equation}\label{eq:modeassumptions}
    \langle h_m|h_m\rangle>1,\quad|\langle h_m|h_{m'}\rangle|\ll1,\quad\langle\epsilon|\epsilon\rangle\ll1,
\end{equation}
where $\langle\cdot|\cdot\rangle$ is the usual detector-noise-weighted inner product on the space of fixed-length time series \cite{Cutler:1994ys}. We will use the term ``modes'' to refer to any set $\{h_m\}$ satisfying \eqref{eq:modedecomp} and \eqref{eq:modeassumptions}, rather than the familiar angular modes $\{h_{lm}\}$ of gravitational radiation. (In fact, $\{h_{lm}\}$ generally does not satisfy the first two conditions of \eqref{eq:modeassumptions}.)

The decomposition \eqref{eq:modedecomp} and \eqref{eq:modeassumptions} can still be achieved phenomenologically for comparable-mass binary mergers by a simple partition of their signals in the frequency domain \cite{Allen:2004gu}, and thus the proposed function \eqref{eq:function} might be beneficial (but not nearly as crucial) for the analysis of such sources. Eqs \eqref{eq:modedecomp} and \eqref{eq:modeassumptions} might also be satisfied by the near-monochromatic signals from quasi-circular early-inspiral binaries, through a partition in the time domain. A more physically motivated decomposition can be obtained for generically inclined and eccentric EMRIs into a Kerr black hole, where the signal power is spread across various harmonics of the three fundamental frequencies; indeed, the canonical modeling approach relies on an angular and frequency-based decomposition $h=\sum h_{lmkn}$ in the first place \cite{Hughes:2021exa}.

Let us now quickly revisit various familiar quantities and identities in GW data analysis, before introducing several new ones. The optimal signal-to-noise ratio (SNR) for a template $h$ is its norm with respect to $\langle\cdot|\cdot\rangle$:
\begin{equation}\label{eq:optsnr}
    \rho(\theta):=\sqrt{\langle h(\theta)|h(\theta)\rangle}.
\end{equation}
We will denote the time-series data from the detector as $x$, and work primarily with two scenarios (hypotheses): the null $H_0$, where $x$ equals the (assumed stationary and Gaussian) detector noise $n$; and $H_1$, where $x=h^*+n$ for a single EMRI signal $h^*$. Further assume that $h^*$ lies in the signal space described by the template model $h(\theta)$, such that $h^*=h(\theta^*)$ for some $\theta^*$. The detection SNR for a template $h$ is the scalar projection of $x$ on $h$, and is a statistic of the data (so we will henceforth refer to it as the standard detection statistic). It is defined as
\begin{equation}\label{eq:detstat}
    X(\theta):=\langle x|\hat{h}(\theta)\rangle:=\frac{\langle x|h(\theta)\rangle}{\rho(\theta)}.
\end{equation}

Recall that for all time series $a,b$, the noise-weighted inner product satisfies the identities
\begin{equation}\label{eq:noiseid}
    \mathrm{E}[\langle n|a\rangle]=0,\quad\mathrm{E}[\langle n|a\rangle\langle n|b\rangle]=\langle a|b\rangle.
\end{equation}
Thus at the template parameters $\theta^*$ (with corresponding optimal SNR $\rho^*$), we have
\begin{equation}\label{eq:detstatdist}
    X|H_0\sim\mathcal{N}(0,1),\quad X|H_1\sim\mathcal{N}(\rho^*,1).
\end{equation}

Finally, we have the standard Bayesian (log-)likelihood function of the source parameters $\theta$, given $H_1$:
\begin{align}\label{eq:loglike}
    \ln{L}(\theta):=&-\frac{1}{2}\langle x-h(\theta)|x-h(\theta)\rangle\nonumber\\
    =\,&\langle x|h\rangle-\frac{1}{2}(\langle h|h\rangle+\langle x|x\rangle).
\end{align}
There is a clearly a close relationship between $X$ and $L$; we can make this more intuitive by observing that the term $\langle x|x\rangle$ is constant over parameter space, while we also have $\partial_\theta\langle h|h\rangle\ll\partial_\theta\langle x|h\rangle$. Thus $\exp{(X\rho)}=\exp{\langle x|h\rangle}$ is an excellent local approximation to $L$ as a density function (i.e., modulo a normalization factor).

We may now define analogous vector-valued versions of $\rho$ and $X$ using the mode decomposition \eqref{eq:modedecomp}, and denote these using boldface by $\boldsymbol{\rho}$ and $\mathbf{X}$. These $M$-vectors are given component-wise by
\begin{equation}\label{eq:optsnrvec}
    [\boldsymbol{\rho}(\theta)]_m:=\rho_m(\theta):=\sqrt{\langle h_m(\theta)|h_m(\theta)\rangle},
\end{equation}
\begin{equation}\label{eq:detstatvec}
    [\mathbf{X}(\theta)]_m:=X_m(\theta):=\langle x|\hat{h}_m(\theta)\rangle:=\frac{\langle x|h_m(\theta)\rangle}{\rho_m(\theta)}.
\end{equation}
It is straightforward to see that
\begin{equation}\label{eq:vecrelation1}
    \mathbf{X}\cdot\boldsymbol{\rho}=\langle x|h\rangle=X\rho
\end{equation}
and, from the second assumption in Eq.\,\eqref{eq:modeassumptions}, that
\begin{equation}\label{eq:vecrelation2}
    |\boldsymbol{\rho}|^2=\boldsymbol{\rho}\cdot\boldsymbol{\rho}=\langle h|h\rangle=\rho^2,
\end{equation}
but note that $|\mathbf{X}|^2=\mathbf{X}\cdot\mathbf{X}\neq\langle x|x\rangle$.

From the noise identities \eqref{eq:noiseid} (and the second assumption in Eq.\,\eqref{eq:modeassumptions}), we have for $\theta^*$ and all $m$:
\begin{equation}\label{eq:detstatvecdist}
    X_m|H_0\sim\mathcal{N}(0,1),\quad X_m|H_1\sim\mathcal{N}(\rho_m^*,1),
\end{equation}
where the $\rho_m^*$ sum to $\rho$ in quadrature from Eq.\,\eqref{eq:vecrelation2}. Thus the chi-squared-like statistic
\begin{equation}\label{eq:chisq}
    \chi^2(\theta):=|\mathbf{X}(\theta)-\boldsymbol{\rho}(\theta)|^2
\end{equation}
has a chi-squared distribution with $M$ degrees of freedom, but only given $H_1$ and when evaluated at the point $\theta^*$.

The appearance of $\chi^2$ in the exponential factor of $f$ suppresses secondary peaks arising from non-local parameter degeneracy, which lifts the main barrier to EMRI search and inference. At this point, the observant reader might note that the exponential amplification of \emph{any} discrepancy between the signal and the secondary template also fulfills the same purpose. Crucially, however, the additional information provided by the effective decomposition of the detection statistic $X$ into $M>1$ modes allows secondary suppression to be achieved without a severe impact on the overall detection sensitivity of $f$ at $\theta^*$. We will return to this point shortly in Sec.\,\ref{subsec:otherfuncs}.

Let us now turn to the remaining undefined quantity in Eq.\,\eqref{eq:function}, which is the calibration factor $\beta$. Without loss of generality, we may sort the decomposition in Eq.\,\eqref{eq:modedecomp} by imposing $\langle h_m|h_m\rangle\leq\langle h_{m+1}|h_{m+1}\rangle$, such that the dominant mode is $h_M$ (which is often $h_{2200}$, in large regions of the signal space). One of the key findings in Chua \& Cutler \cite{Chua:2021aah} is that the strongest secondary peaks in the likelihood (or detection statistic) over the model parameter space are overwhelmingly caused by the dominant mode of a non-local template being phase-aligned with the dominant mode of the actual signal, \emph{without significant contribution from any alignment of the other modes}.

We may use this observation to calibrate the secondary suppression of Eq.\,\eqref{eq:function} with $n=0$ against the noise properties of $X|H_0$; this can be performed using only information about the signal space. For each $\theta$, let us consider the corresponding putative signal $h(\theta)$, and then require that $f^2=\mathrm{Var}(X)=1$ for any putative template with a dominant mode that perfectly matches that of $h(\theta)$, but no other matched modes. Such templates are not explicitly required, of course, since they satisfy by definition
\begin{equation}\label{eq:dommodedetstat}
    X_m=\rho_m(\theta)\delta_{mM},
\end{equation}
where $\delta$ is the Kronecker delta function. For these templates, it follows that
\begin{equation}\label{eq:betaconstrain}
    f=\rho_M(\theta)\exp{\left(-\frac{1}{2}\beta(\theta)(\rho^2(\theta)-\rho_M^2(\theta))\right)}=1.
\end{equation}

Thus we arrive at
\begin{equation}\label{eq:calibration}
    \beta(\theta):=\frac{2\ln{(\alpha(\theta)\rho(\theta))}}{(1-\alpha(\theta)^2)\rho(\theta)^2},
\end{equation}
where we have defined $\alpha:=\rho_M/\rho$ as the fractional optimal SNR of the dominant mode relative to the full template at each parameter point (and so $\alpha^2$ is the fractional power). Note that $\beta(\alpha)$ is monotonically increasing; also,
\begin{equation}\label{eq:alphabounds}
    1<M<\rho^2,\quad\frac{1}{\sqrt{M}}<\alpha<\frac{\sqrt{\rho^2-M+1}}{\rho}
\end{equation}
from the first assumption in Eq.\,\eqref{eq:modeassumptions}, so $\beta$ is well defined:
\begin{equation}\label{eq:betabounds}
    \frac{(\ln{\rho^2}-\ln{M})M}{\rho^2(M-1)}<\beta<\frac{\ln{(\rho^2-M+1)}}{M-1}.
\end{equation}

Finally, this particular definition of $\beta$ allows Eq.\,\eqref{eq:function} to be written alternatively as
\begin{equation}\label{eq:functiondirect}
    f=X(\alpha\rho)^{-\chi^2/((1-\alpha^2)\rho^2)},
\end{equation}
which is useful for direct evaluation---but we shall keep working with the exponential form for analytical convenience. The end result of the calibration is, essentially, that any secondary peaks in $f|H_1$ arising from non-local degeneracy will have a maximum height comparable to the typical variation of $X|H_0$ due to detector noise.

\subsection{Relationship to other functions}
\label{subsec:otherfuncs}

It is useful to construe the proposed function $f$ as an approximation to a more easily interpreted function
\begin{equation}\label{eq:functionchisq}
    f'(\theta):=X(\theta)\left(1-\gamma\left(\frac{M}{2},\frac{\beta'(\theta)\chi^2(\theta)}{2}\right)\right),
\end{equation}
where $\gamma$ denotes the regularized lower incomplete gamma function (i.e., the cumulative distribution function of $\beta'\chi^2$ if it were chi-squared-distributed). The approximation of $f'$ by $f$ is exact in the case $M=2$ and $\beta'=\beta$. For $M>2$, the approximation is still reasonably preserved by the calibration factor $\beta'$, which is now $\neq\beta$ but similarly determined by requiring $f'=1$ at the location of a pure dominant-mode secondary in the case of $n=0$:
\begin{equation}\label{eq:calibrationchisq}
    \beta'=\frac{2}{(1-\alpha^2)\rho^2}\gamma^{-1}\left(\frac{M}{2},1-\frac{1}{\alpha\rho}\right).
\end{equation}
The flatter profile in the vicinity of $\theta^*$ of the $1-\gamma$ factor relative to the exponential factor in Eq.\,\eqref{eq:function} means that $f'$ is locally a closer approximation to $X$ than $f$, which could be beneficial in practical terms. Nevertheless, the function $f'$ is more difficult than $f$ to characterize analytically, and is thus left for future investigation.

When $\beta'$ is simply set to unity across the signal space, $f'$ evokes the chi-squared-weighted SNR seen in ground-based pipelines (as introduced in \cite{LIGOScientific:2011jth}). There, however, $X$ is weighted by an empirically determined function of $\chi^2$ to suppress non-stationary/non-Gaussian noise, and the product of the two has not been fundamentally characterized in the manner of Sec.\,\ref{sec:properties} (to the best of my knowledge). The $\beta'=1$ case is also equivalent to the modified $\langle x|h\rangle$ term in the first of the two ``veto likelihoods'' proposed in Chua \& Cutler \cite{Chua:2021aah}, modulo a factor of $\rho$. The functions in that paper also fulfill the specific aim of secondary suppression, but in a more ad hoc way and, crucially, as an adjustment to $\ln{L}$ rather than $X$.

There exists another interesting interpretation of $f$ as the product of $X$ and $L'$, where the (scaled) logarithm of the latter is written in a deliberately evocative way as
\begin{equation}\label{eq:expfactor}
    \frac{\ln{L'}}{\beta}:=\langle x|h\rangle-\frac{1}{2}\left(\langle h|h\rangle+\sum_m\langle x|\hat{h}_m\rangle\langle\hat{h}_m|x\rangle\right).
\end{equation}
The function $L'$ bears obvious structural similarities to $L$, with the key difference being that the squared norm (with respect to $\langle\cdot|\cdot\rangle$) of $x$ is replaced with the squared norm (with respect to the Euclidean inner product) of its projection onto the mode basis describing each template. One may then wonder whether the function $XL$, or even $XL'$ with $M=1$, might provide a similar suppression of secondaries with suitable calibration. However, there is no way to define a sufficiently discriminative calibration without mode information---any attempt leads to an over-suppression of templates in the local vicinity of $\theta^*$, which is further amplified in the presence of noise and destroys the sensitivity of the template at $\theta^*$ itself.

\section{Properties}
\label{sec:properties}

To study the various geometrical and statistical properties of the function $f$ for EMRI data analysis, we will make use of the AAK template model for generically inclined and eccentric Kerr EMRIs \cite{Chua:2015mua,Chua:2017ujo}, along with the long-wavelength LISA response \cite{Cutler:1997ta}. We keep to the slightly older implementation employed in Chua \& Cutler \cite{Chua:2021aah}, rather than the latest version introduced in \cite{Katz:2021yft} (to avoid redoing the computationally tedious search for secondaries in the space of a new model). As in \cite{Chua:2021aah}, our analysis is restricted to six intrinsic parameters: the two component masses, the spin of the primary mass, and the three initial conditions for the orbit/frequencies. These are commonly parametrized as
\begin{equation}\label{eq:emripars}
    \theta:=\left(\lg{\frac{\mu}{M_\odot}},\lg{\frac{M}{M_\odot}},\frac{a}{M},\frac{p_0}{M},e_0,\cos{\iota_0}\right),
\end{equation}
where $(\mu,M\gg\mu)$ are the masses, $a$ is the usual spin length scale describing Kerr spacetime, and $(p_0,e_0,\iota_0)$ are the initial quasi-Keplerian semi-latus rectum, eccentricity and inclination of the orbit.

The signal injection and ``secondary B'' from \cite{Chua:2021aah} are chosen here as a representative injection and secondary template, respectively; their associated parameters are
\begin{equation}\label{eq:injpars}
    \theta^*=(1,6,0.5,9.5,0.2,0.866),
\end{equation}
\begin{equation}\label{eq:secpars}
    \theta^S=\theta^*+(3,-12,-54,170,1,64)\times10^{-3}.
\end{equation}
While the secondary parameters might not look particularly non-local to the uninitiated reader at first glance, they lie \emph{well beyond} any realistic Bayesian credible region for the inferred parameters. At a typical SNR of 20, the 1-sigma values for these parameters (relative to the maximum a posteriori estimate) are
\begin{equation}\label{eq:bulkpars}
    \pm(0.07,0.2,0.5,3,0.06,0.8)\times10^{-3}.
\end{equation}

There are however some key changes between the setup here and in \cite{Chua:2021aah}. For the present analysis, we consider only full templates sampled at $0.1\,\mathrm{Hz}$, and not the phase-trajectory ``templates'' adopted in that work. White noise and the Euclidean inner product on time series are used here for convenience, and without much loss of generality---this simplification has negligible impact on the correlation structure of signal space at the considered scales, while the properties of any statistics derived from the noise-weighted inner product are preserved. More crucially, the exact definition of modes is slightly different in the two studies. In \cite{Chua:2021aah}, only $M=4$ radial modes with $(l,m,k)=(2,2,0)$ and $-1\leq n\leq2$ are considered; here, we still use $M=4$, $l=2$ and $-1\leq n\leq2$, but with an implicit sum over $(m,k)$ (i.e., the modes now include their sidebands from Lense--Thirring precession, and essentially form the partially decomposed quadrupolar harmonic basis introduced by Barack \& Cutler \cite{Barack:2003fp}). Finally, the signals we examine here are shortened from a duration of two years to six months, but renormalized accordingly to retain an injection SNR of 20. This does not change the location of strong secondaries, at least not significantly (see Sec.\,IV\,B\,1 in \cite{Chua:2021aah} to understand why).

\subsection{As a detection statistic}
\label{subsec:detection}

Let us now examine the statistical properties of $f(\theta^*)$ under the two hypotheses $H_0:x=n$ and $H_1:x=h^*+n$. Closed-form descriptions of the full probability distributions are not straightforward to obtain, but their first couple of moments are comfortably within reach:
\begin{equation}\label{eq:mu0}
    \mathrm{E}[f|H_0]=\frac{\beta\rho}{(\beta+1)^{(M+2)/2}}\exp{\left(-\frac{\beta\rho^2}{2(\beta+1)}\right)},
\end{equation}
\begin{equation}\label{eq:sigma0}
    \mathrm{E}[f^2|H_0]=\frac{4\beta^2\rho^2+2\beta+1}{(2\beta+1)^{(M+4)/2}}\exp{\left(-\frac{\beta\rho^2}{2\beta+1}\right)},
\end{equation}
\begin{equation}\label{eq:mu1}
    \mathrm{E}[f|H_1]=\frac{\rho}{(\beta+1)^{M/2}},
\end{equation}
\begin{equation}\label{eq:sigma1}
    \mathrm{E}[f^2|H_1]=\frac{(2\beta+1)\rho^2+1}{(2\beta+1)^{(M+2)/2}}.
\end{equation}

\begin{figure}[!tbp]
\centering
\includegraphics[width=\columnwidth]{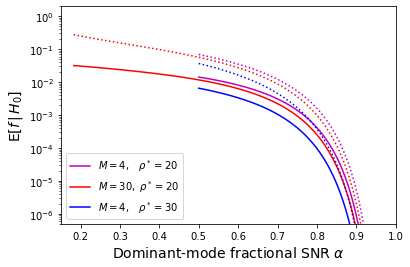}
\includegraphics[width=0.9725\columnwidth,right]{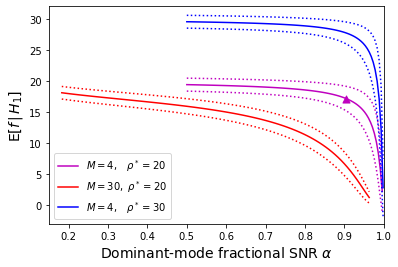}
\caption{Top: Dependence of the means (solid lines) and 1-sigma values (dotted lines) of $f(\theta^*)|H_0$ on $\alpha$, for various $M$ and $\rho^*$. Note that the vertical axis is in logarithmic scale. Bottom: Dependence of the means and 1-sigma values of $f(\theta^*)|H_1$ on $\alpha$. The triangle corresponds to the representative injection considered in this work, for which $\alpha\approx0.9$.}
\label{fig:detection1}
\end{figure}

The full distributions of $f|H_0$ and $f|H_1$ are strongly non-Gaussian, and so the moments \eqref{eq:mu0}--\eqref{eq:sigma1} are only loosely indicative of their location and scale. Nevertheless, the usual means $\mathrm{E}[f]$ and 1-sigma values $\mathrm{E}[f]\pm(\mathrm{E}[f^2]-\mathrm{E}[f]^2)^{1/2}$ are convenient quantities for examining how these distributions depend on $\alpha$, $M$ and $\rho^*$. In Fig.\,\ref{fig:detection1}, the mean and spread of $f|H_0$ and $f|H_1$ are plotted as a function of $\alpha$ for the representative values of $M=4$ and $\rho^*=20$. Also included are curves for an increased number of modes ($M=30$ is approximately the minimum number to describe EMRIs of all eccentricities in the Barack--Cutler harmonic basis \cite{Barack:2003fp}), as well as for a slightly higher injection SNR of 30.

For all considered $M$ and $\rho^*$, the mean and spread of $f|H_0$ are $\ll1$ for most values of $\alpha$---i.e., the statistic $f$ is highly insensitive to noise. Its sensitivity to an actual signal is also reduced from that of $X$, although this reduction is limited for $M=4$ and $\alpha\lesssim0.8$ (since $f|H_1$ retains a mean of $\approx\rho^*$ with a spread of $\approx1$ in that regime). As $\alpha$ approaches its maximal value in Eq.\,\eqref{eq:alphabounds}, the sensitivity of $f$ vanishes, which diminishes its utility for near-equatorial Kerr EMRIs with eccentricities that are $\ll1$ (since $h\approx h_M=h_{2200}$). This is a narrow region of signal space, however, and higher $l$-modes might still carry enough power for $\alpha$ to be suppressed in parts of the region (if sufficient SNR is accumulated close to plunge).

Even with the Barack--Cutler mode decomposition, a fairly large number of modes must be included in Eq.\,\eqref{eq:modedecomp} for a \emph{global} description of the signal space \cite{Barack:2003fp}. This is detrimental to the sensitivity of $f$ (see red curve in Fig.\,\ref{fig:detection1})---less so for high-eccentricity sources, where $\alpha$ is lower due to the larger spread of power across the modes, but certainly for low-eccentricity ones, which would be better analyzed with fewer modes. However, as $M$ is a discrete parameter, we will prefer to avoid varying it across signal space. It is clear from Eq.\,\eqref{eq:mu1} and Fig.\,\ref{fig:detection1} that the ideal decomposition for each signal in the space is with both $\alpha$ and $M$ as small as possible. This might be accomplished for fixed $M$ through a $\theta$-dependent definition of \emph{the decomposition itself}, e.g., relaxing the third assumption in Eq.\,\eqref{eq:modeassumptions} by just choosing the $M$ strongest modes for each template, or even treating $\epsilon$ (the sum of the remaining modes) as a single mode if it satisfies the first two assumptions and is not dominant. Such variants ensure that $f$ still changes smoothly over the signal space; we will leave their study to future global analyses.

\begin{figure}[!tbp]
\centering
\includegraphics[width=\columnwidth]{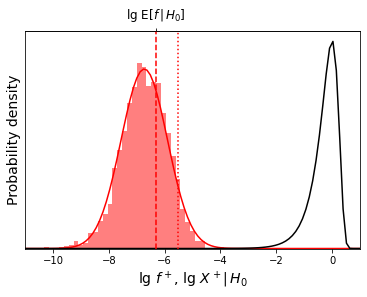}
\includegraphics[width=\columnwidth]{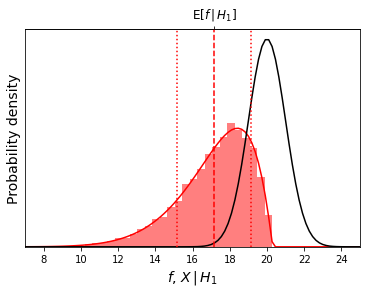}
\caption{Top: Probability distributions of $\lg{f^+(\theta^*)}|H_0$ (red; empirical and normal fit) and $\lg{X^+(\theta^*)}|H_0$ (black) for the representative injection considered in this work. Bottom: Probability distributions of $f(\theta^*)|H_1$ (red; empirical and generalized-beta fit) and $X(\theta^*)|H_1$ (black). In both plots, the exact means (dashed lines) and 1-sigma values (dotted lines) of $f(\theta^*)$ are included for reference.}
\label{fig:detection2}
\end{figure}

To characterize the discriminative power of $f$ as a detection statistic for the representative signal injection considered in this work, we may empirically estimate $p(f|H_0)$ and $p(f|H_1)$ by generating $10^4$ realizations of each statistic, then fitting the samples to a large family of common distributions with a Kolmogorov--Smirnov test. The empirical distribution for $f|H_0$ admits no satisfactory fits, but is observed to satisfy $\mathrm{P}(f>0|H_0)\approx1/2$ to a very good approximation. Since we are concerned only with the positive realizations of $f|H_0$ (denoted by $f^+|H_0$) when computing false-alarm thresholds, we may fit only these and simply reduce the fitted probability density by a factor of two. The distribution of $f^+|H_0$ is best fit by a log-normal distribution with a Kolmogorov--Smirnov statistic $D_{10^4}\approx0.01$, while $f|H_1$ obeys a generalized beta distribution with $D_{10^4}\approx0.006$ (see Fig.\,\ref{fig:detection2}).

With the fitted probability densities $p(f^+|H_0)$ and $p(f|H_1)$ in hand, we may then estimate the false-alarm probability for $f$ given some fixed detection probability. To achieve a minimum detection probability
\begin{equation}\label{eq:sensitivity}
    \mathrm{P}_D:=\int_{f_T}^\infty df\,p(f|H_1)=1-10^{-3},
\end{equation}
the maximum threshold value for $f$ is $f_T\approx9$, and the corresponding false-alarm probability is
\begin{equation}\label{eq:falsealarm}
    \mathrm{P}_F:=\int_{f_T}^\infty df^+\,\frac{1}{2}p(f^+|H_0)\sim10^{-21}.
\end{equation}
An analogous calculation for the standard detection statistic $X$ yields a threshold $X_T\approx17$ and $\mathrm{P}_F\sim10^{-64}$. Thus $f$ is \emph{formally} less discriminative than $X$ due to both its reduced sensitivity and the heavier tail of its (fitted) null distribution, although the practical difference between the two is negligible at an SNR of 20 (and this representative choice of $\theta^*$).

Across the full signal space, non-local degeneracy greatly complicates any attempt at global analysis, and diminishes the validity of idealized treatments, e.g., highly conservative estimates of the required detection-statistic threshold at fixed false-alarm probability for a putative bank of uncorrelated templates $\{h(\theta_i)\}$ spanning the space \cite{Chua:2017ujo,Moore:2019pke}. Furthermore, the joint distributions of the corresponding statistics $\{f_i\}$ under both hypotheses depend strongly on the mode structure across signal space, as opposed to those for $\{X_i\}$. The single-template analysis from above would seem to indicate that $\{f_i\}$ is more prone to Type-I\!I errors than $\{X_i\}$ for a fixed false-alarm probability; however, it also completely fails to address global issues such as the increased Type-I error rate of $\{X_i\}$ over $\{f_i\}$ when detecting the exact set of signals present in data, since it cannot account for false alarms due to secondaries of actual signals.

\subsection{As an objective function}
\label{subsec:identification}

The identification aspect of search is in essence an optimization problem---specifically, the global maximization of some objective function that describes how well the data matches each point in the space of a template model. Pure optimization techniques are highly efficient when maximizing concave(-down) or even log-concave functions, while optimization through sampling is more suitable for functions that are only concave or log-concave beyond a bounded region \cite{2018arXiv181108413M}. Sampling, or at least stochastic optimization, would seem to be the only viable option for non-textbook objective functions such as $X$ and $L$. Although these functions are log-concave in a highly localized region, their large-scale structure is not (unless a globally log-concave prior is specified, but this rather defeats the purpose of search). Furthermore, they suffer not only from intrinsic variations due to non-local signal correlations (with EMRIs being the most extreme manifestation of this), but also from noise variations that are locally correlated on the same length scales as the points in the template model.

\begin{figure}[!tbp]
\centering
\includegraphics[width=\columnwidth]{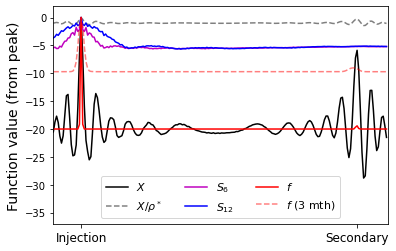}
\includegraphics[width=\columnwidth]{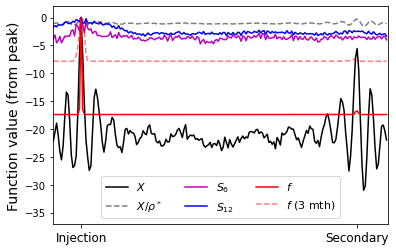}
\caption{Top: Extended connections of $X$, $S_6$, $S_{12}$ and $f$ between injection and secondary parameters, without noise ($x=h^*$). Also included are an annealed standard detection statistic $X/\rho^*$, as well as $f$ for a different signal space (with a truncated duration of three months from the reference initial time). Bottom: As above, but with noise ($x=h^*+n$).}
\label{fig:identification1}
\end{figure}

While the function $f$ is derived with secondary suppression as the primary motivation, it has the beneficial side effect of significantly suppressing noise as well; this fact turns out to be quite useful for the purposes of identification. In Fig.\,\ref{fig:identification1}, the values of $f|H_1$ and $X|H_1$ are compared along an (extended) connecting line in parameter space between the injection parameters $\theta^*$ and the considered secondary parameters $\theta^S$. Following Chua \& Cutler \cite{Chua:2021aah}, we will refer to the restriction of each function to this one-dimensional domain as a \emph{connection}. Not shown in the figure is $\ln{L}$, which is effectively $X\rho^*$ with a slight gradient in the baseline due to changes in $\rho(\theta)$ (these become significant on global length scales, and so the likelihood is ill suited to search). Alternatively, $\exp{X}$ is locally an annealed version of $L$ with temperature $\rho^*$.

Also included in the comparison is a semi-coherent statistic $S_N$ where both data and template are transformed into the time--frequency domain (with a time resolution of $N$), and the statistic is the sum of $N$ \emph{less informative} statistics defined on the individual time segments. This concept is most simply realized through a short-time discrete Fourier transform of both time series, with a partition into $N$ similar-length segments (indexed by $i$) and no window. We may define a semi-coherent inner product between data and template as 
\begin{equation}\label{eq:semicoherentinner}
    \langle x|h\rangle_N:=\sum_{i=1}^N\max_\mathrm{ext.}{\langle x_i|h_i\rangle},
\end{equation}
where the maximization is over a subset of the extrinsic degrees of freedom. These are not the extrinsic parameters per se, and can bleed into the intrinsic parameters in the standard EMRI parametrization. Note that $\langle x|h\rangle_1\neq\langle x|h\rangle$; also, for white noise (or whitened time series with the Euclidean inner product), each $\langle x|h\rangle_N$ reduces exactly to $\langle x|h\rangle$  in the absence of maximization.

For present purposes, we will restrict to analytically maximizing over an overall time shift \cite{blair1991detection}, but not an overall phase shift \cite{Damour:1997ub} for all modes (which is less trivial to perform analytically in the case of EMRIs). Our semi-coherent statistic itself is then defined as
\begin{equation}\label{eq:semicoherentstat}
    S_N:=\frac{\langle x|h\rangle_N}{\sqrt{\langle h|h\rangle_N}}=\frac{\langle x|h\rangle_N}{\rho}=\langle x|\hat{h}\rangle_N.
\end{equation}
Maximizing over some extrinsic degrees of freedom naturally simplifies the global structure of $S_N$, at the cost of losing information about those degrees of freedom. This manifests as a significantly broadened peak for $S_N$ around $\theta^*$, where the broadening increases with $N$ (see Fig.\,\ref{fig:identification1}). More generally, semi-coherent filtering for EMRI search \cite{Gair:2004iv} is an oft-discussed but as yet unactualized paradigm, since it provides an effective means of constructing objective functions with similar properties to $S_N$. It is also highly suited to the nature of LISA data, being overtly compatible with estimation of the non-stationary detector noise \cite{Edwards:2020tlp,Cornish:2020odn}, as well as analysis approaches in the presence of data gaps \cite{Baghi:2019eqo,Dey:2021dem}. While promising in principle, however, we will see that the main benefit of semi-coherent statistics such as $S_N$ (a broadened peak around $\theta^*$) is slightly negated by the retained susceptibility to noise (which is on par with that of $X$).

Rather than performing multiple realizations of full-scale Markov-chain Monte Carlo simulations, we may gain some insight into the \emph{intrinsic traversability} of various objective functions by simply examining the typical speed of a directed Metropolis--Hastings chain along the connecting line from secondary to injection. This approach is used to conduct a ceteris-paribus comparison of the noisy ($x=h^*+n$) functions shown in the bottom panel of Fig.\,\ref{fig:identification1}, with minimal and identical assumptions on the sampling algorithm that is used. It is clear that $X$ itself is significantly more difficult to traverse than the other functions, so we shall anneal it by an effective temperature of $\rho^*$ to ensure a fair fight. Other functions have also been proposed in the literature (or are currently under investigation), both generally and specifically for EMRIs \cite{Chua:2021aah,Jaranowski:1998qm,Wang:2014ica}. We do not consider these here, but only note that the same set of criticisms raised in this manuscript are likely to apply (they are either overly ad hoc, or susceptible to noise).

In this simple directed-chain analysis, a Markov chain is constrained to lie on the connecting line $\{\theta_i\}$ from $\theta^S$ ($i=1$) to $\theta^*$, with a fixed-step transition $\pm(\theta_{i+1}-\theta_i)$ that is small enough to resolve the fine structure in all connections (i.e., the variations due to noise). The proposal distribution is
\begin{equation}\label{eq:proposal}
    \mathrm{P}_P(\theta_j|\theta_i)=
    \begin{cases}
    1-1/(G+1) & j>i \\
    1/(G+1) & i<j 
    \end{cases},
\end{equation}
where $G>1$ is a ``diffusion gradient'' of sorts, and the acceptance probability for each objective function $F$ is
\begin{equation}\label{eq:acceptance}
    \mathrm{P}_A(\theta_j|\theta_i;F)=\min{\left\{1,\frac{1}{G}\exp{(F(\theta_j)-F(\theta_i))}\right\}}.
\end{equation}

\begin{figure}[!tbp]
\centering
\includegraphics[width=\columnwidth]{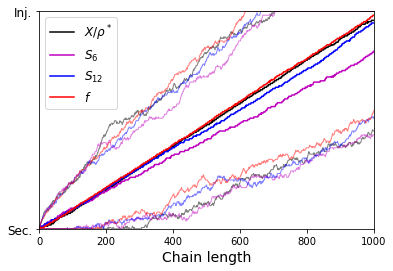}
\caption{Minimum, median and maximum locations of $10^3$ chains for each objective function (with maximum length $10^3$), directed along the connecting line from secondary to injection. The diffusion gradient is $G=2$, chosen such that approximately half of the chains for $f$ reach the injection parameters.}
\label{fig:identification2}
\end{figure}

Each Markov chain is allowed a maximum length of $10^3$, and $10^3$ such chains are produced for the four functions $X/\rho^*$, $S_6$, $S_{12}$ and $f$. The results are summarized in Fig.\,\ref{fig:identification2}, where the minimum, median and maximum locations of the chain (i.e., the indices of $\theta_i$) are shown at each step in the chain. Perhaps unsurprisingly, the median chain speed for each objective function is dictated by the degree of noise in the corresponding connections; the least-traversable function is $S_6$, followed by $S_{12}$ (which is visually smoother from the bottom panel of Fig.\,\ref{fig:identification1}), then $X/\rho^*$ (which retains the secondary peak but is even smoother in the tail region) and finally $f$ (which is virtually constant in the tail region). This trend does not appear to depend on the diffusion gradient, with the speed order preserved at various examined values of $G$.

The above analysis is not fully conclusive, of course, since the chains are directed along a line towards the global peak for the sake of efficiency. In a higher-dimensional setting, a search with $f$ will take longer to chance upon the peak, whose extent \emph{at the baseline value} is determined by the local length scale of fully coherent signal correlations (the same holds for $X$ and $\ln{L}$, regardless of any annealing factor). This is not nearly as localized as the \emph{posterior bulk} itself, whose characteristic length scale is inversely proportional to the injection SNR; see Sec.\,\ref{subsec:inference}. Thus the global peak in $f$ is no more difficult to find than its counterpart in, say, the highest-temperature likelihood of a parallel-tempering algorithm. On the other hand, while the larger ``capture region'' of the semi-coherent statistic will no doubt be beneficial, strong non-local variations due to signal correlations are still present in its tail region (this is not evident along the connections shown in Fig.\,\ref{fig:identification1}, but the baseline is certainly not $\approx-5$ across the signal space). Whether or not these larger-scale gradients will be a net help or hindrance to semi-coherent search remains to be determined.

It is also possible to construct $f$-based functions with broadened peaks by conducting less informative comparisons between data and template, but without resorting to the partial maximization or marginalization of the function over some degrees of freedom. The most direct approach is simply to alter the signal space altogether, by truncating the signal templates (and data) to some shorter observation duration with the original reference initial time (defined implicitly by $(p_0,e_0,\iota_0)$). An example of $f$ for a truncated duration of three months is shown in Fig.\,\ref{fig:identification1} (dashed red connections), where the broadening of the primary peak and the retained suppression of all other variations are both evident. The function $f$ with a range of truncated observation durations might thus be used in a sampling scheme that is akin to parallel tempering, or similar hierarchical-style searches.


We have thus far been considering only the single-signal hypothesis $H_1$ for identification. In the case of multiple EMRI signals, the difficulty in searching the signal space is amplified for all functions. However, it is not hard to see that the complexity of $f$ will only be weakly affected (unless secondaries from different signals are likely to combine, but an analytical calculation in Chua \& Cutler \cite{Chua:2021aah} indicates that they are not). Strong peaks in $f$ exist only at the parameters of actual signals; while search chains will tend to be trapped at each location, this does not detract from either parallelized or sequential search, and is even beneficial as it facilitates a direct transition into inference (see Sec.\,\ref{subsec:inference}).

The efficiency of sampling algorithms when used for optimization relies not only on the global structure of the objective function, but also the intrinsic cost of evaluating the function. While the evaluation of the vector-valued detection statistic $\mathbf{X}$ incurs an overt factor of $M$ in computational complexity over $X$ (or $L$), this only multiplies the cost of the inner-product operation $\langle\cdot|\cdot\rangle$. The cost of each such operation certainly dominates the total cost on the signal-processing end---but the EMRI template $h$ itself is also notoriously expensive to generate without computational enhancements, and ultimately even irreducibly so due to its sheer length and complexity. Depending on the efficiency of the template model, the function $f$ should thus be evaluable at little to modest additional computational cost over $X$ or $L$.

\subsection{As a likelihood function}
\label{subsec:inference}

Finally we turn to the procedure of inference, by examining the utility of $f|H_1$ as an \emph{effective} (log-)likelihood function. Recall that at leading order, the standard log-likelihood \eqref{eq:loglike} in the vicinity of $\theta^*$ is simply \cite{Vallisneri:2007ev}
\begin{equation}\label{eq:leadloglike}
    \ln{L}=-\frac{1}{2}\Delta^T\mathcal{I}\Delta+\mathcal{O}(|\Delta|^3),
\end{equation}
where $\Delta:=\theta-\theta^*$, and the Fisher information $\mathcal{I}$ is given component-wise by the pullback metric on signal space:
\begin{equation}\label{eq:fisher}
    \mathcal{I}=\langle\partial_\theta h|\partial_\theta h\rangle|_{\theta=\theta^*}.
\end{equation}
At a sufficiently high SNR, the full likelihood becomes well represented by Eq.\,\eqref{eq:leadloglike}, and the resultant posterior appears close to Gaussian in the bulk region. For the sake of concreteness, we will define the posterior-bulk region here as the set of points where $\ln{L}$ exceeds the 3-sigma value $-9/2$ for the leading-order likelihood (with a locally uninformative prior).

Assuming the components of $\partial_\theta\rho^2=2\langle h|\partial_\theta h\rangle$ and $\partial_\theta\beta$ are negligible in the bulk (which is generally valid for all parameters but the luminosity distance), a similar expansion of $f$ about $\theta^*$ yields
\begin{equation}\label{eq:leadf}
    f=-\frac{1}{2}\Delta^T\mathcal{I}'\Delta+\mathcal{O}(|\Delta|^3),
\end{equation}
where
\begin{equation}\label{eq:fisherf}
    \mathcal{I}'=\beta\rho\left(\mathcal{I}-\sum_m\mathcal{I}_m\right)+\frac{1}{\rho}\mathcal{I},
\end{equation}
with the ``Fisher information'' for each mode given by
\begin{equation}\label{eq:fishermode}
    \mathcal{I}_m=\langle\partial_\theta h_m|\partial_\theta h_m\rangle|_{\theta=\theta^*}.
\end{equation}
The exponential factor in Eq.\,\eqref{eq:function} contributes to the first term in Eq.\,\eqref{eq:fisherf}, while the second term is proportional to the Fisher information and arises from the expansion of $X$ about $\theta^*$ (with the additional factor of $1/\rho$ due to the template normalization in $X$).

Positive-definiteness induces a partial ordering on the set of all symmetric matrices (and thus the subset of positive-definite matrices as well). We now assume that $\mathcal{I}_m\prec\mathcal{I}_{m+1}\prec\mathcal{I}$ for all $m$. Then (very roughly):
\begin{equation}\label{eq:partialorderapp}
    \mathcal{I}-\sum_m\mathcal{I}_m\approx\mathcal{I}-\mathcal{I}_M\approx(1-\alpha^2)\mathcal{I},
\end{equation}
and we have
\begin{equation}\label{eq:fisherfapp}
    \mathcal{I}'\approx\frac{2\ln{(\alpha\rho)}+1}{\rho}\mathcal{I}.
\end{equation}

\begin{figure}[!tbp]
\centering
\includegraphics[width=\columnwidth]{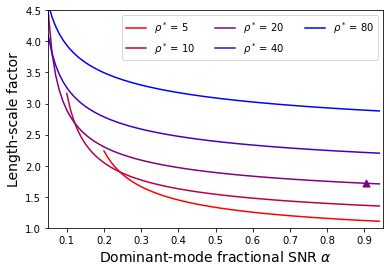}
\caption{Dependence of the estimated length-scale factor \eqref{eq:lengthscalefactor} on $\alpha$, for various $\rho^*$. The triangle corresponds to the representative injection considered in this work, for which $\alpha\approx0.9$.}
\label{fig:inference1}
\end{figure}

Since the decomposition \eqref{eq:modedecomp} for EMRIs is in terms of signal-like modes with the same underlying dependence on the source parameters, these modes vary in qualitatively similar ways to the full signal template, such that Eq.\,\eqref{eq:fisherfapp} is not a terrible approximation. In fact, the local profile of $f$ about $\theta^*$ generally looks consistent with negative excess kurtosis, in which case the approximation is actually conservative. For $\rho>5$ and all $\alpha$, Eq.\,\eqref{eq:fisherfapp} indicates a broadening of $f$ over $\ln{L}$, since we have
\begin{equation}\label{eq:lengthscalefactor}
    \sqrt{\frac{\rho}{2\ln{(\alpha\rho)}+1}}>1,
\end{equation}
where the left-hand side is the associated length-scale factor. This quantity is plotted as a function of $\alpha$ in Fig.\,\ref{fig:inference1}, for various values of $\rho^*$ (it does not depend on $M$ due to the crude approximation in Eq.\,\eqref{eq:partialorderapp}).

\begin{figure}[!tbp]
\centering
\includegraphics[width=\columnwidth]{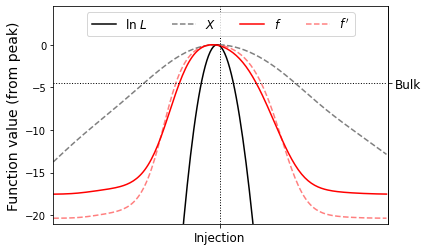}
\caption{Extended connections of $\ln{L}$, $X$, $f$ and $f'$ between injection and secondary parameters, with noise and zoomed in on (a one-dimensional slice of) the posterior-bulk region.}
\label{fig:inference2}
\end{figure}

For our representative signal injection, $\alpha\approx0.9$ and the estimated length-scale factor is $\approx1.7$---this is a slight underestimate of the actual broadening of $f$ over $\ln{L}$, which is by a factor of between two and three (see Fig.\,\ref{fig:inference2}). Also included in Fig.\,\ref{fig:inference2} are the profiles of $f'$ from Eq.\,\eqref{eq:functionchisq} (which, recall, is slightly more flat-topped and locally closer to $X$), as well as $X$ itself (where the broadening over $\ln{L}$ is by a factor of $(\rho^*)^{1/2}\approx4.5$, as expected).

With the general broadening of $f$ over $\ln{L}$, the extent of the recovered distribution from the sampling of $f$ provides useful prior localization for regular posterior estimation with $L$. More promisingly, note from Fig.\,\ref{fig:inference1} that the broadening is also modest (no more than a factor of five at an SNR of 20). This constrained broadening of $f$ over $\ln{L}$ implies that the search chains might (upon a suitable level of convergence) be used \emph{directly as posterior samples}, with the unnormalized importance weights
\begin{equation}\label{eq:weights}
    w:=L\pi\exp{(-f)},
\end{equation}
where $\pi$ is the desired Bayesian prior.

\section{Conclusion}
\label{sec:conclusion}

In this manuscript, I propose a potentially useful one-stop function $f(\theta)$ for various tasks in GW data analysis; it is conditioned on the detector data $x$, and defined for some signal-template model $h(\theta)$ with a general mode decomposition. The basic mechanics of $f$ rely critically on a set of assumptions about the decomposition---essentially that signal templates may readily be written as the sum of $M>1$ uncorrelated modes, each with substantial power. This work is primarily motivated by the deep-rooted problem of \emph{strong non-local parameter degeneracy} in the space of EMRI signals, and so recent results by Chua \& Cutler \cite{Chua:2021aah} on the nature of this phenomenon are used to define a model-specified calibration of the function for EMRI data analysis. I then build a case for the utility of $f$ by examining its properties as a statistic for detection, as an objective function for identification, and as an effective likelihood function for inference.

The main difficulty in EMRI data analysis is the procedure of source identification, which is essentially a large-scale optimization problem. Traditionally defined objective functions over the model space, such as the standard detection statistic or standard likelihood function in GW data analysis, suffer from the presence of numerous and highly pronounced secondary peaks that hinder identification. The function proposed here is based on the principle of \emph{de-emphasizing} these peaks in some way, which is shared by well-known strategies such as semi-coherent filtering or annealing-type sampling algorithms. Semi-coherent filtering also falls under another class of strategies (including, e.g., $F$-statistic searches \cite{Jaranowski:1998qm}) that use partial maximization or marginalization to conduct less informative comparisons between data and template. This is useful for EMRI identification in another way, since it broadens all peaks in the function---including the primary peak containing the global maximum, which is highly localized relative to the model space.

I posit a different tenet here: that a function with virtually no gradients is easier to search than one with uncontrolled variations at both large and small length scales. This is realized in the proposed function by using \emph{exponential suppression} to de-emphasize secondaries, rather than congealing them (as in the case of semi-coherent filtering) or simply rescaling the entire function (annealing). As a by-product, any less-pronounced variations in the function due to signal correlations or detector noise are suppressed to a near-constant baseline. I argue in Sec.\,\ref{subsec:identification} that the elimination of such variations might play a larger role in the efficacy of search algorithms than any other individual factor, although the jury is still out on whether exponential suppression is generally superior to partial maximization or vice versa. To provide a useful (if somewhat biased) analogy, the latter is akin to searching for a bigger needle in a haystack (really a field of haystacks), while the former only involves finding a smaller needle in an open field.

The ``one-stop'' aspect of the proposed function also connects identification to detection and inference more seamlessly, at least at the conceptual level. In practice, of course, one must still assess the statistical significance of signal candidates through traditional methods, and will likely take the effort to repeat inference on identified signals directly with the standard likelihood---but neither is intrinsically challenging to do for EMRIs. As briefly discussed in Sec.\,\ref{subsec:identification}, the function can easily be used in hierarchical approaches to replicate the advantages of partial maximization. It is also viable in the broader context of the LISA global fit: search and inference for individual (resolvable) EMRI signals will almost certainly be performed independently using the catalog residuals, as they do not strongly impact noise estimates, or inference on signals of a different source type \cite{Racine:2007gv} (or even on one another \cite{Chua:2021aah}). That being said, some thought might be required to adapt the function for the presence of data gaps. Beyond the setting of EMRI data analysis, the general principle of exponential suppression should also, at minimum, provide an interesting alternative to the existing paradigms in coherent GW searches.

\acknowledgements

I am grateful to Curt Cutler for seminal discussions, and for our closely related collaboration on \cite{Chua:2021aah}. Lorenzo Speri suggested the effective broadening of the proposed function with truncated observation durations, for which I am deeply appreciative. Thanks also go out to Katerina Chatziioannou, Neil Cornish, Michael Katz, Christopher Moore and Michele Vallisneri for their insightful input on the manuscript. This work was supported by the NASA LISA Preparatory Science grant 20-LPS20-0005.

\bibliography{main}

\end{document}